%% file: main.tex
\documentclass[sigconf]{acmart}
\usepackage{amsmath}
\usepackage{pifont}
\usepackage{algorithm}      
\usepackage{algpseudocode}  
\usepackage{caption}    
\usepackage{graphicx}
\usepackage{siunitx}
\usepackage{booktabs,tabularx,makecell,threeparttable}
\usepackage{mathtools}
\usepackage{tikz}
\usepackage{multirow}
\usetikzlibrary{fit,calc}

\AtBeginDocument{%
  }

\setcopyright{acmlicensed}
\copyrightyear{2018}
\acmYear{2018}
\acmDOI{XXXXXXX.XXXXXXX}
\acmConference[Conference acronym 'XX]{Make sure to enter the correct
  conference title from your rights confirmation email}{June 03--05,
  2018}{Woodstock, NY}

\acmISBN{978-1-4503-XXXX-X/2018/06}

\begin{document}

\title{Towards Reliable Social A/B Testing: Spillover-Contained Clustering with Robust Post-Experiment Analysis}

\author{Xu Min, Zhaoxu Yang, Kaixuan Tan, Juan Yan, Xunbin Xiong, Zihao Zhu\\
        Kaiyu Zhu, Fenglin Cui, Yang Yang, Sihua Yang, Jianhui Bu}

\affiliation{%
  \institution{Kuaishou Technology}
  \city{Beijing}
  \country{China}
}

\email{{minxu, yangzhaoxu, tankaixuan, yanjua03, xiongxunbin, zhuzihao03}@kuaishou.com}
\email{{zhukaiyu, cuifenglin03, yangyang35, yangsihua, bujianhui}@kuaishou.com}

\renewcommand{\shortauthors}{Min et al.}

\begin{abstract}

A/B testing is the foundation of decision-making in online platforms, yet social products often suffer from network interference: user interactions cause treatment effects to spill over into the control group. Such spillovers bias causal estimates and undermine experimental conclusions.
Existing approaches face key limitations: user-level randomization ignores network structure, while cluster-based methods often rely on general-purpose clustering that is not tailored for spillover containment and has difficulty balancing unbiasedness and statistical power at scale.
We propose a spillover-contained experimentation framework with two stages. In the pre-experiment stage, we build social interaction graphs and introduce a \emph{Balanced Louvain} algorithm that produces stable, size-balanced clusters while minimizing cross-cluster edges, enabling reliable cluster-based randomization. In the post-experiment stage, we develop a tailored CUPAC estimator that leverages pre-experiment behavioral covariates to reduce the variance induced by cluster-level assignment, thereby improving statistical power. Together, these components provide both structural spillover containment and robust statistical inference.
We validate our approach through large-scale social sharing experiments on Kuaishou, a platform serving hundreds of millions of users.
Results show that our method substantially reduces spillover and yields more accurate assessments of social strategies than traditional user-level designs, establishing a reliable and scalable framework for networked A/B testing.
\end{abstract}



\received{20 February 2007}
\received[revised]{12 March 2009}
\received[accepted]{5 June 2009}

\maketitle

\input{sections/01_introduction}

\input{sections/02_related_work}

\input{sections/03_method}

\input{sections/04_experiments}

\input{sections/06_conclusion}

\bibliographystyle{ACM-Reference-Format}
\bibliography{ref}

\appendix

\input{sections/appendix}

\end{document}

%% file: sections/01_introduction.tex
\section{Introduction}

Online controlled experiments (commonly known as A/B testing) have become a fundamental tool for data-driven decision-making in modern internet platforms. 
A key identification assumption underlying A/B testing is the Stable Unit Treatment Value Assumption (SUTVA), which requires that each unit’s outcome depends only on its own treatment assignment and is unaffected by the treatment status of others \cite{rubin1980randomization,rubin1986comment,rubin2005causal}. 
While SUTVA is often plausible in independent user scenarios, it is routinely violated in social networks due to \emph{network interference} \cite{hudgens2008toward, manski2013identification}.

\begin{figure}[t]
    \centering
    \includegraphics[width=0.8\linewidth]{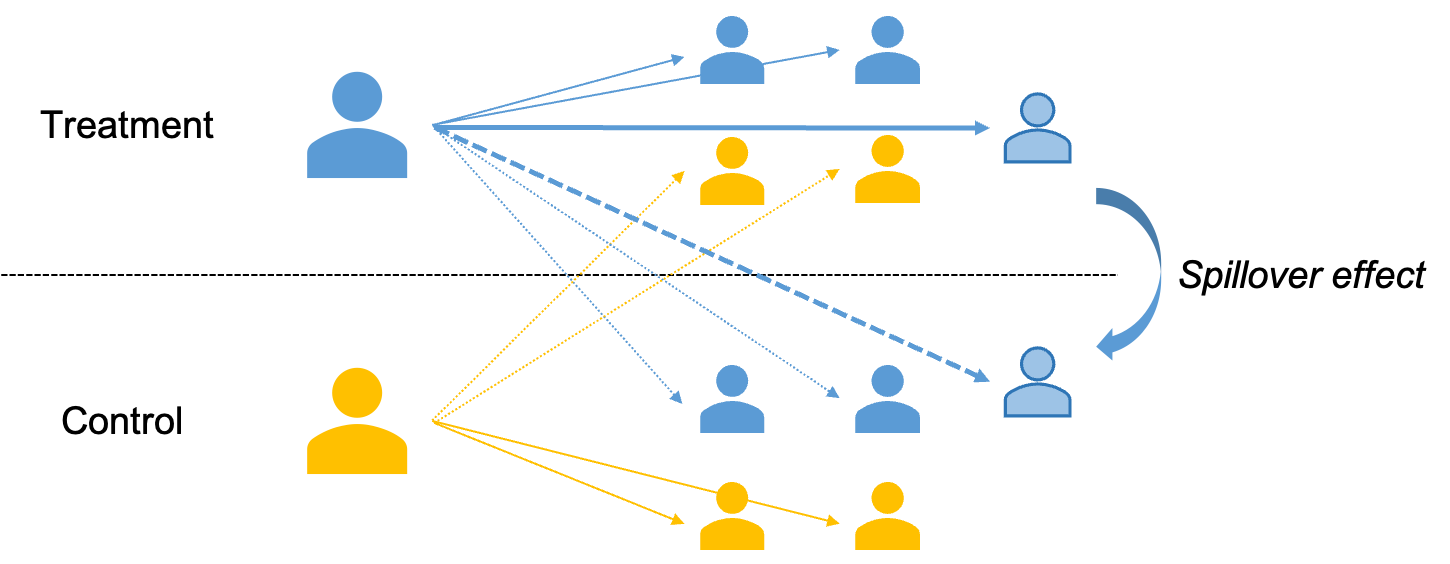}
    \caption{Spillover effects in social A/B testing: treated users influence control users through interactions (e.g., sharing), biasing effect estimates.}
    \label{fig:spillover}
\end{figure}

In social platforms such as Kuaishou, TikTok, or Facebook, users are densely connected through sharing, messaging, and following relationships. 
As illustrated in Figure \ref{fig:spillover}, treatment effects can spill over to untreated users via interactions, violating SUTVA and biasing estimates \cite{aronow2017estimating,tchetgen2012causal,gui2015network}.
This leads to (i) estimation bias and (ii) user experience risks from inconsistent exposure among connected peers.

Existing research on interference in experimentation generally falls into two complementary categories.  
First, \textbf{randomization-based designs} reduce spillovers through improved assignment \cite{eckles2017design, cai2023independent, ugander2013graph}. 
Examples include switchback testing \cite{bojinov2023design}, geographic isolation \cite{vaver2011measuring}, cluster-based randomization \cite{eckles2017design, holtz2025reducing, ugander2023randomized}, and two-sided designs \cite{karrer2021network, nandy2021b}. 
However, they can reduce effective sample size and require stable, balanced clusters.

Second, \textbf{post-experiment analysis} methods relax SUTVA assumptions and attempt to correct interference effects after data collection. 
Techniques include adjusting estimates based on network structure, filtering out potentially contaminated units, and employing advanced estimators that model exposure conditions \cite{aronow2017estimating, savje2021average, bhadra2025causal, farias2023correcting}. 
While these approaches preserve sample efficiency and adapt flexibly to dynamic networks, they rely heavily on modeling assumptions and may lack interpretability or robustness in business-critical settings.

For Kuaishou, where network-driven spillover effects are first-order, stability and reliability of inference are paramount.
To this end, we present a production-ready experimentation framework that integrates \emph{cluster-based randomization} with \emph{variance-reduction} techniques for industrial-scale social A/B testing.
Our approach starts by constructing a business-driven social interaction graph that encodes heterogeneous relationships (e.g., sharing, messaging, following).
We then introduce \emph{Balanced Louvain}, a clustering algorithm tailored to experimentation, which explicitly targets (i) low cross-cluster connectivity to contain spillovers and (ii) balanced cluster sizes to support stable, scalable randomization.
Finally, to mitigate the variance inflation inherent in cluster-level assignment, we employ CUPAC, leveraging pre-experiment covariates to improve sensitivity without sacrificing the spillover-containment benefits.
We deploy this framework on Kuaishou and validate its effectiveness across multiple social experiments; in a sharing-strategy experiment, it substantially reduced spillover contamination and produced more reliable, actionable effect estimates than conventional user-level randomization.

Our contributions are summarized as follows:
\begin{itemize}
    \item We propose \emph{Balanced Louvain}, a clustering approach tailored to randomized experiments, optimizing cluster stability, size balance, and spillover containment for large-scale social networks.
    \item We develop a post-experiment analysis pipeline for cluster-based experiments that leverages CUPAC for variance reduction, improving statistical power under cluster randomization.
    \item We deploy and validate the proposed framework on Kuaishou, demonstrating robustness and practical value in real-world social experiments.
\end{itemize}

%% file: sections/02_related_work.tex
\section{Related work}

\subsection{Overall Designs for Networked Experiments}
A large body of research has investigated methods to mitigate network interference in A/B testing, which can be broadly categorized into design-based approaches and post-experiment analytical approaches.

\begin{figure*}[t]
    \centering
    \includegraphics[width=0.8\textwidth]{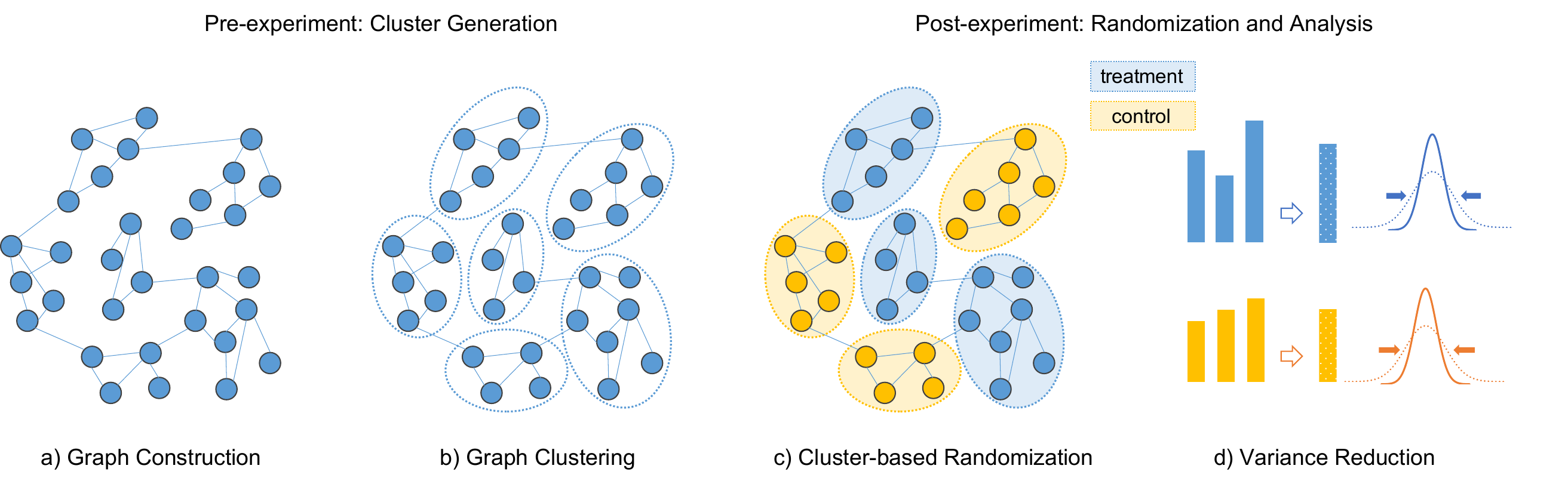}
    \caption{Overview of our proposed framework for reliable social A/B testing.
    The system consists of two main stages: (1) pre-experiment graph construction and clustering to minimize cross-cluster spillovers;
    (2) post-experiment cluster-based randomization and effect estimation with variance reduction (CUPAC) for more sensitive detection.}
    \label{fig:overview}
\end{figure*}

\subsubsection{Design-based approaches}
Design-based methods aim to reduce spillover by carefully controlling treatment assignment ex-ante. 
Switchback testing \cite{bojinov2023design} alternates treatment assignment across time windows, mitigating temporal interference but being less effective when user interactions span across time. 
Geographic randomization \cite{vaver2011measuring} assigns treatment at the level of geographical units (e.g., cities, delivery zones), ensuring interference is contained spatially, though it often sacrifices granularity and statistical power.
Independent-set designs \cite{cai2023independent} instead identify non-interfering subsets of users within the graph and randomize treatment within these sets, offering unbiasedness but often at the cost of discarding many units.
In addition, two-sided or bipartite experiments \cite{karrer2021network,nandy2021b} are widely used in industrial practice, especially in platform scenarios such as content sharing, where treatment is separately applied and evaluated on the sender and receiver sides.
Cluster-based randomization \cite{eckles2017design, ugander2013graph, holtz2025reducing, ugander2023randomized} partitions the social graph into clusters, ensuring that interactions primarily remain within treatment groups. 
While effective in mitigating spillover, clustering reduces the effective sample size and places strong requirements on cluster stability.
Additionally, existing clustering methods lack explicit optimization for A/B test objectives, focusing instead on general community detection rather than minimizing cross-cluster spillover.

\subsubsection{Post-experiment analysis approaches}
An alternative line of work relaxes the SUTVA assumption and instead corrects for interference during analysis. 
For instance, exposure re-weighting estimators \cite{aronow2017estimating,savje2021average} explicitly model the probability that a unit experiences interference given network connections. 
Recent advances propose interference-aware causal estimators that account for network structure to debias estimates \cite{ma2021causal,farias2023correcting, yuan2021causal}.
More recently, graph-based adjustment methods \cite{bhadra2025causal} incorporate structural information into estimation, improving robustness to heterogeneous spillover patterns.
Analytical approaches preserve statistical power but rely on strong assumptions about interference structure, making them sensitive to model mis-specification and harder to interpret in production settings.

\subsection{Graph Clustering Algorithm}
Conventional clustering algorithms such as K-means or DBSCAN \cite{mcqueen1967some, ester1996density} operate in a Euclidean feature space, grouping points based on vector similarity.
In social A/B testing, however, the core problem is \emph{graph partitioning}: we seek clusters that keep interactions within experimental units to contain spillovers.
Graph-based clustering algorithms exploit network topology.
Connected component clustering \cite{fortunato2010community} groups nodes by reachability, while Label Propagation Algorithm (LPA) \cite{raghavan2007near} iteratively assigns communities based on neighbor labels.
Louvain \cite{blondel2008fast} and its successor Leiden \cite{traag2019louvain} are widely used at scale, primarily optimizing modularity.

While modularity-oriented methods often yield high intra-community density, they are not explicitly designed for experimentation requirements such as (i) controlling cross-cluster edges as a direct proxy for spillover exposure, (ii) producing \emph{balanced} and operationally manageable cluster sizes for randomization, and (iii) maintaining stability under graph dynamics.
These gaps motivate our Balanced Louvain algorithm, which adapts Louvain-style optimization to better align clustering objectives with spillover containment and size balance in industrial A/B testing.

\subsection{Variance Reduction in A/B Testing}
Statistical power is a crucial consideration in online experiments, especially when interference-aware designs (e.g., clustering) reduce the effective number of independent randomization units.
To mitigate variance inflation, covariate-adjusted estimators have been widely studied.
CUPED \cite{deng2013improving} uses pre-experiment covariates correlated with the outcome to form a control variate, reducing variance while maintaining unbiasedness.
Building on this idea, CUPAC \cite{tang2020control} incorporates richer auxiliary covariates produced by predictive models and achieves additional variance reduction beyond CUPED.

Variance reduction is particularly important for cluster-randomized experiments: clustering can substantially reduce spillover but typically increases estimator variance due to cluster-level assignment.
However, most CUPED/CUPAC formulations are developed under individual-level randomization and do not directly capture the variance structure of cluster-split experiments \cite{lin2024variance}.
While prior work \cite{wang2024model} discusses cluster-randomized settings, existing studies rarely provide an end-to-end framework that jointly (i) constructs spillover-contained clusters and (ii) couples the design with a practical variance-reduction estimator.
Our work bridges this gap by combining Balanced Louvain clustering in the design stage with a tailored CUPAC-based post-experiment analysis pipeline.

%% file: sections/03_method.tex
\section{Methodology}
\label{sec:methodology}

This section presents an end-to-end framework for reliable social A/B testing under network interference.
The key idea is a two-stage design: we \emph{contain spillovers by design} through experimentation-oriented clustering (Balanced Louvain on a multi-behavior interaction graph), and we \emph{recover statistical efficiency in analysis} through a CUPAC-based inference pipeline tailored to cluster randomization.
We first introduce the overall framework, then describe our Balanced Louvain clustering algorithm (spillover containment, size balance, and temporal stability), and finally present the bucket-based estimation and CUPAC adjustment procedure.

\subsection{Overall Framework}
\label{sec:framework}

We propose a two-stage framework tailored for large-scale social experiments, combining pre-experiment spillover containment with post-experiment variance reduction (Figure~\ref{fig:overview}).

\textbf{Stage I (design): spillover-contained experimental units.}
We construct a multi-behavior interaction graph and run Balanced Louvain to produce clusters that are (i) \emph{spillover-contained} (few cross-cluster edges), (ii) \emph{size-balanced} (operationally safe traffic splitting), and (iii) \emph{stable} over time.
These properties make cluster-level randomization practical and reproducible in production.

\textbf{Stage II (analysis): power recovery with CUPAC.}
Cluster-level assignment reduces the effective number of randomization units and can inflate variance.
We therefore adopt a CUPAC-based estimator that leverages pre-treatment behavioral covariates to improve precision while preserving the interpretability of treatment effects.

Together, the framework yields a reliable workflow for social A/B testing: we mitigate spillovers before deployment and recover statistical efficiency after deployment.

\subsection{Balanced Graph Clustering Algorithm}

\subsubsection{Key challenges}

Let $G = (V, E, W)$ denote a weighted social network, where $V$ is the set of $n$ users, $E$ is the set of edges representing social connections, and $W: E \rightarrow \mathbb{R}^+$ represents edge weights.
In social A/B testing, clustering is not only a community-detection task: the resulting clusters become \emph{experimental units} for randomization.
Accordingly, we aim to partition $V$ into $k$ disjoint clusters $\mathcal{C} = \{C_1, C_2, \ldots, C_k\}$ that satisfy three experimentation-driven requirements:
\begin{itemize}
    \item \textbf{Spillover containment} – minimize cross-cluster connections to reduce interference between treatment groups.
    \item \textbf{Size balance} – control cluster sizes to support operationally safe traffic splitting and stable variance.
    \item \textbf{Temporal stability} – maintain consistent cluster assignments across experiment periods.
\end{itemize}

\subsubsection{Balanced Louvain Algorithm}
\label{sec:balanced_louvain}

We propose a \textbf{Balanced Louvain Algorithm} that tailors the standard Louvain procedure to the needs of large-scale social experimentation.
The core insight is that modularity optimization alone may produce extremely large communities, which are undesirable for cluster-level randomization.
Balanced Louvain therefore augments Louvain with explicit \emph{size control} while preserving its scalability and its tendency to keep dense interactions within clusters---a key driver of spillover containment.
Concretely, we combine a soft size-aware objective during optimization with a hard size constraint in post-processing, producing clusters that are both experimentation-ready and scalable.

\paragraph{Standard Modularity Optimization}
The standard Louvain algorithm optimizes modularity $Q$, defined as:
\begin{equation}
    Q = \frac{1}{2m} \sum_{i,j} \left[ w_{ij} - \gamma \frac{k_i k_j}{2m} \right] \delta(c_i, c_j)
\end{equation}
where $m = \sum_{ij} w_{ij} / 2$ is the total edge weight, $k_i = \sum_j w_{ij}$ is the weighted degree of node $i$, $\gamma$ is the resolution parameter, and $\delta(c_i, c_j) = 1$ if nodes $i$ and $j$ belong to the same cluster.

When considering moving node $i$ from its current cluster to cluster $C$, the modularity gain is:
\begin{equation}
    \Delta Q_i^C = k_{i,in}^C - \gamma \frac{k_i \cdot \Sigma_{tot}^C}{2m}
    \label{eq:modularity_gain}
\end{equation}
where $k_{i,in}^C$ is the sum of edge weights from node $i$ to nodes in cluster $C$, and $\Sigma_{tot}^C$ is the sum of degrees of all nodes in cluster $C$.

\paragraph{Soft Constraint: Size Penalty}
To discourage the formation of overly large clusters during optimization, we introduce an explicit size penalty. The modified score for moving node $i$ to cluster $C$ becomes:
\begin{equation}
    S_i^C = \Delta Q_i^C - \alpha \cdot P(|C|)
    \label{eq:soft_constraint}
\end{equation}
where $\alpha \geq 0$ is the size balance factor and $P(|C|)$ is the size penalty function.

We design $P(|C|)$ as a piecewise function that only penalizes clusters exceeding a threshold:
\begin{equation}
    P(|C|) =
    \begin{cases}
        0 & \text{if } |C| \leq \tau \\
        \bar{k} \cdot \frac{|C| - \tau}{\tau} & \text{if } |C| > \tau
    \end{cases}
    \label{eq:penalty_function}
\end{equation}
where $\tau$ is the penalty threshold (typically set to $\frac{N_{max}}{2}$ where $N_{max}$ is the maximum allowed cluster size), and $\bar{k} = \frac{2m}{n}$ is the average node degree used for normalization.
The choice of $\bar{k}$ as the normalization factor is crucial for ensuring that the penalty term operates at the same scale as the modularity gain. 
Since the typical modularity gain $\Delta Q_i^C$ is approximately $O(\bar{k})$ for an average node, using $\bar{k}$ in the penalty function makes $\alpha$ an interpretable parameter: it directly represents the fraction of modularity gain one is willing to sacrifice for better cluster balance. 
For instance, $\alpha = 0.3$ means the penalty at $|C| = N_{max}$ equals 30\% of a typical modularity gain.

This design ensures:
\begin{itemize}
    \item Small and medium clusters are not penalized, preserving modularity optimization.
    \item Large clusters receive increasing penalties as they grow beyond the threshold.
    \item The penalty magnitude is normalized to match the scale of modularity gains.
\end{itemize}

\paragraph{Hard Constraint: Post-Processing Split}
After the algorithm converges, we apply a hard constraint to split any cluster exceeding the maximum size $N_{max}$. Unlike random bisection, we use a \textbf{connectivity-based intelligent splitting} approach.
The splitting process repeats until all clusters satisfy the size constraint:

For each oversized cluster $C$ with $|C| > N_{max}$:
\begin{enumerate}
    \item Compute internal connectivity for each node $i \in C$:
    \begin{equation}
        \text{conn}_i = \sum_{j \in C, j \neq i} w_{ij}
    \end{equation}
    \item Sort nodes by connectivity in ascending order.
    \item Create a new empty cluster $C_{new}$.
    \item Iteratively move nodes with the lowest connectivity from $C$ to $C_{new}$ until $|C| \leq N_{max}$.
\end{enumerate}

This iterative process continues until no cluster exceeds the size limit. 
The approach minimizes modularity loss by removing peripheral nodes that have weak connections to the cluster core, while the original cluster retains its densely connected core.

\paragraph{Algorithm Description}
Algorithm~\ref{alg:balanced_louvain} presents the Balanced Louvain algorithm, highlighting our key innovations: the soft constraint during node movement and the connectivity-based hard constraint in post-processing.

\begin{algorithm}[t]
\caption{Balanced Louvain Algorithm}
\label{alg:balanced_louvain}
\begin{algorithmic}[1]
\Require Graph $G = (V, E, W)$, balance factor $\alpha$, max size $N_{max}$
\Ensure Cluster assignment $\mathcal{C} = \{C_1, \ldots, C_k\}$

\State Initialize each node as its own cluster; compute $\tau \leftarrow N_{max} / 2$
\Repeat
    \For{each node $i \in V$}
        \For{each neighboring cluster $C$}
            \State Compute modularity gain $\Delta Q_i^C$ using Eq.~\eqref{eq:modularity_gain}
            \State Compute size penalty $P(|C|)$ using Eq.~\eqref{eq:penalty_function}
            \State \textbf{[Soft Constraint]} $S_i^C \leftarrow \Delta Q_i^C - \alpha \cdot P(|C|)$ \hfill $\triangleright$ Eq.~\eqref{eq:soft_constraint}
        \EndFor
        \State Move $i$ to cluster with highest $S_i^C$ if positive
    \EndFor
    \State Contract graph (standard Louvain phase 2)
\Until{convergence}

\State \textbf{[Hard Constraint]} Post-processing:
\Repeat
    \For{each cluster $C$ with $|C| > N_{max}$}
        \State Compute connectivity: $\text{conn}_i = \sum_{j \in C} w_{ij}$ for each $i \in C$
        \State Create new cluster $C_{new} \leftarrow \emptyset$
        \While{$|C| > N_{max}$}
            \State Move node with lowest $\text{conn}_i$ from $C$ to $C_{new}$
        \EndWhile
    \EndFor
\Until{no cluster exceeds $N_{max}$}
\State \Return $\mathcal{C}$
\end{algorithmic}
\end{algorithm}

\paragraph{Parameter Configuration}
Our algorithm introduces three key parameters:

\begin{itemize}
    \item \textbf{Size Balance Factor} ($\alpha$): Controls the strength of the soft constraint. We recommend $\alpha \in [0.2, 0.5]$. Higher values produce more balanced clusters but may reduce modularity.

    \item \textbf{Maximum Cluster Size} ($N_{max}$): The hard upper bound on cluster size. This parameter should be set based on experimental requirements, such as the maximum acceptable sample size imbalance when a cluster is assigned to a treatment group. In practice, we recommend setting $N_{max}$ to limit the largest cluster to a small fraction (e.g., 0.1\%-1\%) of the total population.

    \item \textbf{Resolution Parameter} ($\gamma$): Standard Louvain parameter controlling cluster granularity. We use $\gamma = 1$ as the default.
\end{itemize}

\paragraph{Complexity Analysis}
The time complexity of our algorithm is $O(n \log n)$ per iteration, matching the standard Louvain algorithm. 
The additional size penalty computation adds only $O(1)$ overhead per node movement decision. The post-processing split has complexity $O(|C| \log |C|)$ for each oversized cluster $C$. 
Analysis details can be found in Appendix~\ref{appendix:bl_complexity}.

\paragraph{Theoretical Justification}
Our soft constraint design is motivated by the observation that modularity optimization inherently favors large clusters. 
The penalty term in Eq.~\eqref{eq:penalty_function} counteracts this bias by making large clusters less attractive for new nodes. 
Importantly, by normalizing the penalty with $\bar{k}$, we ensure that the penalty magnitude is comparable to the modularity gain, allowing $\alpha$ to serve as an intuitive balance parameter.

Unlike the global resolution parameter $\gamma$, which fragments \emph{all} communities when increased, our $P(|C|)$ is a \emph{piecewise, local} penalty: it is zero below $\tau$ and only activates when a candidate community becomes too large, preserving small communities while limiting the growth of oversized ones.

The connectivity-based splitting in the hard constraint phase preserves community structure better than random splitting. 
By removing peripheral nodes first, we maintain the core of each community, which typically contains the most densely connected subgraph.

\subsubsection{Multi-Behavior Graph Construction for Stability}
To improve cluster stability, we build a heterogeneous interaction graph that aggregates multiple social behaviors (e.g., follow, comment, like, share, message) instead of relying on a single relation.
We define the edge weight between users $i$ and $j$ as:
\begin{equation}
W_{ij} = \sum_{d=1}^{D} \omega_d \cdot s^{(d)}_{ij},
\end{equation}
where $s^{(d)}_{ij}$ is the interaction strength of behavior type $d$ and $\omega_d$ is its importance weight.

\subsubsection{Engineering Implementation for Scalability}
\label{subsubsec:eng}

Deploying Balanced Louvain on Kuaishou-scale graphs requires careful systems co-design to avoid excessive shuffle and synchronization overheads in distributed graph processing.
We summarize the key optimizations in Appendix~\ref{appendix:scalability}.

\subsection{Evaluation with CUPAC Adjustment}

\subsubsection{Key Challenges}

ATE estimation under cluster-based randomization with CUPAC adjustment faces two key challenges:

\begin{itemize}
    \item \textbf{Accurate ATE Estimation and Inference} – Cluster-based randomization complicates unbiased ATE estimation and valid inference due to intra-cluster correlation.

    \item \textbf{Safe Variance Reduction} – Variance reduction methods, such as CUPAC that leverage predictive modeling, must avoid inflating Type I error to improve precision and test power.
\end{itemize}

\subsubsection{Bucket-Based ATE Estimation under Cluster-Based Randomization}

In cluster-based randomization, users are grouped into clusters, and clusters are randomly assigned to treatment or control. Because users within the same cluster are often correlated, treating them as independent samples can underestimate the variance, leading to overly optimistic significance tests.

To address this challenge, statistical inference is typically conducted at the cluster level or higher. 
In practice, clusters can be hashed into buckets and randomly assigned to treatment or control, and group-level metrics are computed by aggregating within buckets.
This approach preserves approximate independence of units used for inference while correctly accounting for intra-cluster correlation.

Let $Y_b$ denote the aggregated response metric of bucket $b$, and $N_b$ the bucket size. Denote the sets of treatment and control buckets by $\mathcal{B}_{\mathrm{treat}}$ and $\mathcal{B}_{\mathrm{ctrl}}$. The aggregated ratios for treatment and control groups can be written concisely as
\begin{equation}
\bar{R}_{\mathrm{treat}} = \frac{\sum_{b \in \mathcal{B}_{\mathrm{treat}}} Y_b}{\sum_{b \in \mathcal{B}_{\mathrm{treat}}} N_b}, \quad
\bar{R}_{\mathrm{ctrl}} = \frac{\sum_{b \in \mathcal{B}_{\mathrm{ctrl}}} Y_b}{\sum_{b \in \mathcal{B}_{\mathrm{ctrl}}} N_b},
\end{equation}
and the group-level ATE is
\begin{equation}
\hat{\tau}_{\mathrm{agg}} = \bar{R}_{\mathrm{treat}} - \bar{R}_{\mathrm{ctrl}}.
\end{equation}

\paragraph{Bucket-Level Linearization for Ratio Metrics.}
A first-order Delta-method \cite{hosseini2019unbiased} is applied to ratio metrics $R_b = Y_b / N_b$ to define a bucket-level pseudo-outcome:
\begin{equation}
Z_b = \frac{\mu_Y}{\mu_N} + \frac{1}{\mu_N} Y_b - \frac{\mu_Y}{\mu_N^2} N_b,
\end{equation}
where $\mu_Y$ and $\mu_N$ are the reference means of $Y_b$ and $N_b$ computed across all buckets.

This transformation offers several advantages:
\begin{enumerate}\setlength{\itemsep}{0pt}\setlength{\topsep}{2pt}\setlength{\parsep}{0pt}\setlength{\parskip}{0pt}
    \item \textbf{Provides approximate unbiasedness.} 
    The bucket-level pseudo-outcome $Z_b$ provides an approximately unbiased estimate of the aggregated ratio:
    \begin{align}
        \bar{R}_{\mathrm{treat}} &\approx \frac{1}{|\mathcal{B}_{\mathrm{treat}}|} \sum_{b \in \mathcal{B}_{\mathrm{treat}}} Z_b, \\
        \bar{R}_{\mathrm{ctrl}} &\approx \frac{1}{|\mathcal{B}_{\mathrm{ctrl}}|} \sum_{b \in \mathcal{B}_{\mathrm{ctrl}}} Z_b.
    \end{align}
    
    \item \textbf{Direct variance estimation.} 
    The variance of the group-level ratios can be computed directly across buckets:
    \begin{align}
        \mathrm{Var}(\bar{R}_{\mathrm{treat}}) &= \frac{1}{|\mathcal{B}_{\mathrm{treat}}|^2} 
        \sum_{b \in \mathcal{B}_{\mathrm{treat}}} \mathrm{Var}(Z_b), \\
        \mathrm{Var}(\bar{R}_{\mathrm{ctrl}}) &= \frac{1}{|\mathcal{B}_{\mathrm{ctrl}}|^2} 
        \sum_{b \in \mathcal{B}_{\mathrm{ctrl}}} \mathrm{Var}(Z_b),
    \end{align}
    where the variance of each transformed ratio $Z_b$ is estimated directly, without decomposition into $Y_b$ and $N_b$.

    \item \textbf{Facilitates CUPAC modeling.} 
    Using $Z_b$ allows direct bucket-level modeling, simplifying the application of CUPAC while maintaining interpretable ATE and variance estimates for the original ratio metrics.
\end{enumerate}

\paragraph{Inference Using Bucket-Level Estimates.}  
Conditional on bucket-level randomization, the variance of the aggregated ATE is
\begin{equation}
\widehat{\mathrm{Var}}(\hat{\tau}_{\mathrm{agg}}) = \mathrm{Var}(\bar{R}_{\mathrm{treat}}) + \mathrm{Var}(\bar{R}_{\mathrm{ctrl}}).
\end{equation}

The resulting $t$-statistic for testing the null hypothesis of zero treatment effect is
\begin{equation}
t = \frac{\hat{\tau}_{\mathrm{agg}}}{\sqrt{\widehat{\mathrm{Var}}(\hat{\tau}_{\mathrm{agg}})}},
\end{equation}
with degrees of freedom approximated using the bucket-level\\ 
Welch–Satterthwaite equation:
\begin{equation}
\nu = \frac{\left( \mathrm{Var}(\bar{R}_{\mathrm{treat}}) + \mathrm{Var}(\bar{R}_{\mathrm{ctrl}}) \right)^2}
{\frac{\mathrm{Var}(\bar{R}_{\mathrm{treat}})^2}{|\mathcal{B}_{\mathrm{treat}}| - 1} + \frac{\mathrm{Var}(\bar{R}_{\mathrm{ctrl}})^2}{|\mathcal{B}_{\mathrm{ctrl}}| - 1}},
\end{equation}
so that the $t$-statistic approximately follows a Student-$t$ distribution with $\nu$ degrees of freedom. Here, $|\mathcal{B}_{\mathrm{treat}}|$ and $|\mathcal{B}_{\mathrm{ctrl}}|$ denote the number of buckets in the treatment and control groups, respectively.

This bucket-level formulation ensures robust and approximately unbiased ATE estimation while providing correct variance estimation and interpretable inference for ratio metrics under cluster-based randomization.

\subsubsection{Variance Reduction with CUPAC under Clustered Experiments}

Cluster-based randomization often induces positive correlation of outcomes within clusters, inflating the variance of naive ATE estimators. To improve precision, variance reduction methods leveraging pre-treatment information can be applied. In our framework, variance is first reduced using CUPED with a single covariate and further enhanced with CUPAC incorporating multiple covariates or model-based predictions; a simple illustration of the resulting variance reduction is shown in Figure~\ref{fig:overview}(d).

\paragraph{CUPED and CUPAC}  
Classical \emph{CUPED} \cite{deng2013improving} adjusts outcomes using a single covariate $X_b$ to reduce variance:
\begin{equation}
Y_b^{\mathrm{CUPED}} = Y_b - \theta (X_b - \mathbb{E}[X]), 
\quad \theta = \frac{\mathrm{Cov}(Y,X)}{\mathrm{Var}(X)}.
\end{equation}

\emph{CUPAC} \cite{deng2013improving} generalizes this approach by using multiple covariates or machine-learning predictions $\hat{Y}_b$:
\begin{equation}
Y_b^{\mathrm{CUPAC}} = Y_b - \theta (\hat{Y}_b - \mathbb{E}[\hat{Y}]), 
\quad \theta = \frac{\mathrm{Cov}(Y, \hat{Y})}{\mathrm{Var}(\hat{Y})}.
\end{equation}

Both preserve unbiasedness while reducing variance proportionally to the predictive power of the covariates.

\paragraph{Operational Pipeline for CUPAC}  
ATE estimation under cluster-based randomization is conducted at the experiment and bucket level to capture time-specific patterns and within-experiment behavioral variations. Applying CUPAC at this granularity can improve variance reduction and precision relative to global predictors \cite{poyarkov2016boosted}. However, sampling variability and potential overfitting at the bucket level may inflate Type I error rates, which must be carefully controlled during model training and prediction.
To obtain robust, unbiased, and precise ATE estimates while preserving valid hypothesis testing, we adopt the following operational pipeline:

\begin{enumerate}\setlength{\itemsep}{0pt}\setlength{\topsep}{2pt}\setlength{\parsep}{0pt}\setlength{\parskip}{0pt}
    \item \textbf{Delta-Method Transformation.}  
    Compute bucket-level pseudo-outcomes $Z_b$ from the original numerator $Y_b$ and denominator $N_b$. This stabilizes variance and provides suitable inputs for CUPAC modeling.

    \item \textbf{Covariate Selection.}  
    Select bucket-level covariates $\mathbf{X}_b$ pre-experiment, based on historical experiments and prior data, that are both robust and predictive. Robustness is ensured by their stability across past experiments and time periods, while predictive power is assessed via LASSO or feature importance scores. This pre-selection ensures that CUPAC adjustment is effective and reliable, without using post-treatment information.

    \item \textbf{Cross-Fitted Prediction.}  
    Partition control buckets into $K$ folds. For each fold $k$, train a predictive model $f^{(k)}$ on the remaining $K-1$ folds and generate out-of-sample predictions $\hat{Z}_b^{(k)}$ for the held-out fold. Only control buckets are used for training to avoid bias from treatment.  
    For treatment buckets, predictions are averaged across the $K$ models:
    \begin{equation}
        \hat{Z}_b = \frac{1}{K} \sum_{k=1}^{K} f^{(k)}(\mathbf{X}_b), \quad b \in \mathcal{B}_{\mathrm{treat}}.
    \end{equation}
    Cross-fitting mitigates overfitting, reduces sampling variability, and stabilizes Type I error rates \cite{wang2024model}. Empirically, we have observed that using cross-fitting in a decision-tree model reduces Type I error from 8.54\% to 5.08\%, demonstrating its empirical effectiveness.

    \item \textbf{CUPAC Adjustment.}  
    Compute CUPAC-adjusted pseudo-outcomes:
    \begin{equation}
        Z_b^{\mathrm{CUPAC}} = Z_b - \theta (\hat{Z}_b - \mathbb{E}[\hat{Z}]), 
        \quad
        \theta = \frac{\mathrm{Cov}(Z, \hat{Z})}{\mathrm{Var}(\hat{Z})}.
    \end{equation}

    \item \textbf{Adjusted ATE Estimation.}  
    Aggregate the CUPAC-adjusted pseudo-outcomes to obtain the final ATE:
    \begin{equation}
        \hat{\tau}_{\mathrm{CUPAC}} =
        \frac{1}{|\mathcal{B}_{\mathrm{treat}}|} \sum_{b \in \mathcal{B}_{\mathrm{treat}}} Z_b^{\mathrm{CUPAC}}
        - \frac{1}{|\mathcal{B}_{\mathrm{ctrl}}|} \sum_{b \in \mathcal{B}_{\mathrm{ctrl}}} Z_b^{\mathrm{CUPAC}}.
    \end{equation}
\end{enumerate}

This pipeline provides a model-robust, bucket-level covariate adjustment that effectively reduces variance while preserving unbiased ATE estimation and valid hypothesis testing under cluster-based randomization.

%% file: sections/04_experiments.tex
\section{Experiments}

This section evaluates our spillover-contained experimentation framework from both a principled and a practical perspective.
We first use controlled simulations to validate the design rationale, then present online experiments on Kuaishou to describe the experimental setting and report business impact.
Finally, we provide focused analyses of the two key components in our framework: Balanced Louvain for experimentation-oriented clustering and CUPAC for post-experiment variance reduction.

\subsection{Principle Validation via Simulation}

To validate the core design principle---that stronger spillover containment yields less biased effect estimation---we conduct controlled simulations on synthetic social networks.

\textbf{From Clustering to Experiment: The WGSR Metric.}
The intra-cluster edge ratio $\rho(\mathcal{C})$ quantifies clustering quality at the graph level, but it does not directly capture spillover containment in an A/B test.
In our experiments, we use \textit{Within-Group Share Ratio} (WGSR) as the experiment-level spillover metric, defined in Section~\ref{subsubsec:eval_metrics}.
Intuitively, higher WGSR indicates stronger spillover containment.
In well-contained cluster randomization, WGSR increases as $\rho(\mathcal{C})$ increases, whereas under user-level randomization it degrades toward 0.5.

\begin{figure}[t]
\centering
\includegraphics[width=\linewidth]{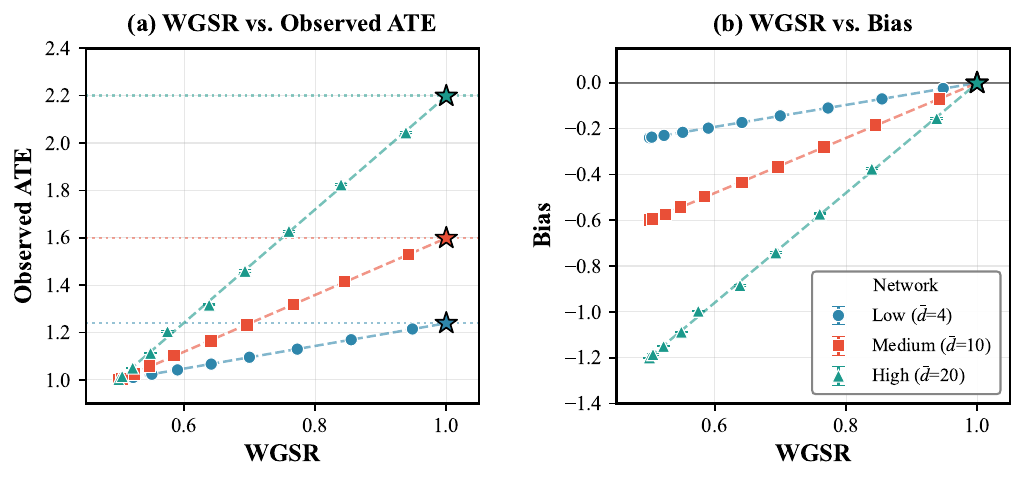}
\caption{WGSR vs.\ observed ATE (left) and bias (right). Cluster-based assignment reduces bias by 87--89\%. Horizontal dotted lines indicate the true ATE. Stars: extrapolation to WGSR$=$1.0 recovers true ATE within 0.1\%.}
\label{fig:wgsr_bias}
\end{figure}

\textbf{Setup and Results.}
We generate Watts-Strogatz networks and vary WGSR from 0.50 to 0.95 (900 experiments total; see Appendix~\ref{appendix:simulation}). 
As shown in Figure~\ref{fig:wgsr_bias}, observed ATE increases linearly with WGSR ($R^2 > 0.999$), and cluster-based assignment (WGSR$\approx$0.95) reduces bias by 87--89\% compared to UID randomization (WGSR$\approx$0.50). 
Higher network density amplifies spillover: bias reaches $-1.2$ for $\bar{d}{=}20$ versus $-0.24$ for $\bar{d}{=}4$ under UID randomization.
High WGSR mitigates estimation bias regardless of density.

\vspace{-6pt}

\subsection{Real-World Experimental Setup}
We conducted large-scale online A/B experiments on Kuaishou's production platform to evaluate the framework under real social interactions.
We describe the application scenario and treatment strategy, the parallel experiment design, and the evaluation system used to quantify both spillover containment and business impact.

\subsubsection{Application Scenario and Treatment Design}
In social platforms, sharing exhibits a strong bilateral network effect: the behavior of the sender and the recipient is inherently interdependent.
As a result, conventional user-level randomization suffers from spillovers---treated users influence control users through sharing pathways---which biases ATE estimates downward (Figure~\ref{fig:spillover}).

To estimate the full social treatment effect, we apply our spillover-contained framework: we partition the interaction graph into clusters and randomize treatment at the cluster level.
The treatment consists of a set of recommender and UI strategies (e.g., optimized video ranking and user icon reordering in the sharing panel) designed to increase users' propensity to share.

\subsubsection{Experimental Design} 
We performed two online A/B experiments that adopted {UID-Based Randomization (UBR)} and {Cluster-Based Randomization (CBR)}, respectively.
All other experimental configurations are identical between the two trials, except for the traffic-splitting mechanism.

To ensure complete isolation and eliminate any possibility of cross-experiment contamination, the allocation schemes for the CBR and UBR experiments are arranged as illustrated in Appendix \ref{appendix:uid-cid-exp}.
This parallel setup allows us to observe the same strategy under different randomization frameworks simultaneously, eliminating temporal fluctuations in treatment effects and thereby improving the confidence of our conclusions.

\subsubsection{Evaluation Metrics}
\label{subsubsec:eval_metrics}
The experiment evaluates the superiority of our approach from two perspectives: network spillover effects and business-oriented metrics.

For the network spillover effect, we define \textbf{Within-Group Share Ratio (WGSR)}.
Specifically, for a given experimental group $g$ with a control group $g^{\prime}$ of equal size, let
  \begin{equation}
  \text{WGSR}_g = \frac{\#\{(u{\to}v)\in E_{\text{share}} \mid u\in g,\, v\in g\}}{\#\{(u{\to}v)\in E_{\text{share}} \mid u\in g, v\in g \cup g^{\prime} \}}.
  \label{eq:wgsr}
  \end{equation}
A higher WGSR under cluster-based diversion directly indicates that the clustering successfully confines social influence inside treatment arms, thereby attenuating network spillover bias.

For the business-oriented metrics, we evaluate the treatment group's relative improvement over the control group by:
\begin{itemize}\setlength{\itemsep}{-1pt}\setlength{\topsep}{0pt}\setlength{\partopsep}{0pt}\setlength{\parsep}{0pt}\setlength{\parskip}{0pt}
  \item \textbf{Share Count (SC)}: total number of sharing events initiated by all users during the observation period.
  \item \textbf{Share Penetration (SP)}: percentage of users who performed at least one sharing action.
  \item \textbf{Message Penetration (MP)}: proportion of users who performed at least one message action (as sharing is one type of message).
  \item \textbf{7-Day Retention (LT-7)}: 7-day retained active users divided by the cohort size, measuring medium-term user stickiness.
\end{itemize}\vspace{-6pt}

\begin{table}[t]
\centering
\small
\setlength{\tabcolsep}{3pt}
\caption{Results of two parallel online A/B test experiments conducted to investigate network effects on a 2\% sample of all Kuaishou users.}
\label{tab:results}
\begin{threeparttable}
\resizebox{\columnwidth}{!}{%
\begin{tabular}{l c c c c c}
\toprule
{Method} & {WGSR(\%)} & {SC(\%)} & {SP(\%)} & {MP(\%)} & {LT-7(\%)} \\
\midrule
{UBR} & 49.68
& +7.758$^{**}$
& +4.968$^{**}$
& +1.888$^{**}$
& +0.049$^{*}$ \\
\textbf{CBR (ours)} & \textbf{90.39}
& \textbf{+26.612}$^{**}$
& \textbf{+14.774}$^{**}$
& \textbf{+4.983}$^{**}$
& \textbf{+0.133}$^{**}$ \\
\bottomrule
\end{tabular}%
}
\begin{tablenotes}
\footnotesize
\item Note: $^{*}$ denotes $p<0.05$ and $^{**}$ denotes $p<0.01$.
\end{tablenotes}
\end{threeparttable}
\end{table}

\subsection{Online Experimental Results}
\subsubsection{Graph Construction and Clustering}

For the monthly active users on Kuaishou, we built an ultra-large weighted social graph to support graph clustering, which incorporated the top five social relationships correlated with sharing behavior. 
Based on this graph, we applied Balanced Louvain ($\alpha=0.3$, $N_{\max}=250\mathrm{K}$) to generate the user clusters for the CBR experiment.

The resulting clustering scheme partitioned approximately 370 million users into 73,340 distinct clusters. 
The largest cluster contains 249,319 users, with an average cluster size of 5,045. 
The overall modularity is 0.466, indicating a well-defined community structure.
Notably, 44.7\% of all sharing events occurred within clusters, demonstrating that the clustering approach effectively mitigates network spillover effects on experimental metrics.

\subsubsection{Results of A/B Testing.}
\label{sec:online_results}
As shown in Table \ref{tab:results}, the CBR design leads to a significant increase in all core metrics compared to the conventional UBR approach, which we attribute to the reduction of network spillover effects.
In particular, receiving a shared item not only increases a user’s consumption but also induces reciprocal sharing, thereby amplifying the gains in both sharing (SC, SP) and messaging (MP).  
This amplification propagates more fully under spillover-contained CBR, yielding higher estimates of downstream social engagement than UBR.  

We further examined the relative gain in 7-day average user retention of the treatment group versus the control group under both UBR and CBR designs.  
The LT-7 effect is statistically insignificant under UBR, whereas the CBR design detects a significant positive LT-7 lift, which is \textbf{0.088pp} higher than that observed under UBR (See Appendix \ref{appendix:lt7}).  
This demonstrates that by controlling spillover, CBR not only clarifies immediate sharing effects but also reveals the long-term retention benefits of sharing strategies.

\subsection{Analysis of Balanced Louvain}
To assess Balanced Louvain as an experimentation-oriented clustering method, we conducted ablation experiments on a city-level subgraph $G_{\mathrm{city}}$ (550K users) from a northwestern Chinese city.

We compare the following methods:
\begin{itemize}
    \item \textbf{LPA (constrained)}: Label Propagation with size constraints (see Appendix \ref{sec:appendix_lpa}).
    \item \textbf{Louvain}: Standard Louvain with different resolutions.
    \item \textbf{Ours}: Balanced Louvain with explicit cluster size control. We report three variants in Table~\ref{tab:quality}: \emph{Hard-only} ($\alpha=0$, $N_{max}=40\mathrm{K}$), \emph{Soft-only} ($\alpha=0.3$, $N_{max}=-40\mathrm{K}$; where negative value disables post-processing split), and \emph{Soft+Hard} ($\alpha=0.3$, $N_{max}=40\mathrm{K}$).
\end{itemize}

\begin{table}[t]
\small
\setlength{\tabcolsep}{2pt}
\centering
\caption{Comparison of clustering methods. Composite score: $\text{Score} = 0.2Q + 0.3\rho + 0.3 \cdot \text{Bal} + 0.2 \cdot \text{Ctrl}$, where $\text{Bal} = 1 - \sigma/\sigma_{max}$ measures size uniformity, and $\text{Ctrl} = 1$ if max cluster is guaranteed within threshold.}
\label{tab:quality}
\resizebox{\columnwidth}{!}{%
\begin{tabular}{lccccccc}
\toprule
{Method} & {Modularity} & {Intra-edge} & {Variance} & {Max Cluster} & {Ctrl} & {Score} \\
\midrule
LPA (constrained) & 0.516 & 0.496 & 742,833 & 44,485 & \checkmark & 0.631 \\
Louvain ($\gamma=1$) & \textbf{0.648} & \textbf{0.548} & 1,839,930 & 109,296 & $\times$ & 0.294 \\
Louvain ($\gamma=2$) & 0.646 & 0.500 & 387,911 & 50,202 & $\times$ & 0.516 \\
\midrule
Ours (Hard only) & 0.613 & 0.496 & 399,240 & 40,000 & \checkmark & 0.707 \\
Ours (Soft only) & 0.623 & 0.511 & 360,540 & 45,722 & $\times$ & 0.519 \\
\textbf{Ours (Soft+Hard)} & 0.622 & 0.511 & 373,842 & 36,877 & \checkmark & \textbf{0.716} \\
\bottomrule
\end{tabular}%
}
\end{table}

Table~\ref{tab:quality} reveals three key findings:
(1) While standard Louvain achieves the highest modularity and intra-edge ratio, it produces extremely high variance (1.84M) and uncontrollable large clusters, resulting in the lowest composite score (0.294).
(2) Our Soft+Hard approach achieves the best composite score (0.716), outperforming LPA (0.631) by 13\% and Hard-only (0.707) by 1\%, with the lowest variance (374K) among controlled methods.
(3) Compared to Louvain ($\gamma=2$), our method achieves better balance (variance 374K vs 388K) while providing precise size control and higher intra-edge ratio (0.511 vs 0.500). 
This confirms that Balanced Louvain successfully finds experimentation-ready communities, whereas general-purpose Louvain creates uncontrollable huge clusters that destabilize variance.

The ablation shows that Soft-only achieves the lowest variance (361K) but cannot guarantee size control (exceeds threshold by 14\%). 
Hard-only guarantees control but has highest variance (399K) among our methods due to disruptive post-processing splits. 
The combination leverages soft constraint to guide balanced community formation while ensuring size guarantees.
Detailed parameter sensitivity experiments are provided in Appendix~\ref{sec:appendix_clustering}.

\subsection{Analysis of CUPAC}

To enhance experimental precision under cluster-level randomization, we implement an operational CUPAC pipeline using stable and predictive pre-treatment covariates.
Measured at the bucket level, these covariates include aggregated user-level social-context metrics (e.g., Share Count) and cluster-level characteristics (e.g., the number of clusters).
These covariates are then used to generate bucket-level predictions via multivariate linear regression, which are subsequently adjusted with CUPAC to produce robust pseudo-outcomes.

\textbf{Variance Reduction Metric.} We quantify the efficiency gain of covariate adjustment relative to the standard Difference-in-Means (DIM) estimator using the variance reduction ratio:
\begin{equation}
\text{VarRed}_{\text{method}} = 1 - \frac{\mathrm{Var}(\hat{\tau}_{\text{method}})}{\mathrm{Var}(\hat{\tau}_{\text{DIM}})},
\end{equation}
where $\hat{\tau}_{\text{method}}$ denotes the ATE estimated using CUPED or CUPAC.

Over the 45-day experimental period, we applied both CUPED and CUPAC for covariate adjustment. 
Notably, CUPAC produced statistically significant gains in Kuaishou LT-7 (main app only) as early as day 28, with narrower confidence intervals compared with DIM and CUPED (Table~\ref{tab:sig_test}). 
This early detection highlights CUPAC’s improved sensitivity in identifying treatment effects. 
Across the full observation period, CUPAC consistently outperforms CUPED, achieving additional variance reduction of approximately 4–7pp for the two LT-7 metrics (see Appendix~\ref{sec:appendB} for detailed results).

\begin{table}[t]
\centering
\small
\caption{Kuaishou LT-7 (day28): comparison across methods}
\label{tab:sig_test}
\begin{tabular}{lccc}
\toprule
{Method} & {DIM} & {CUPED} & \textbf{CUPAC (ours)} \\
\midrule
{Relative diff (\%)} & 0.169 & 0.130 & \textbf{0.122} \\
{P-value}            & 0.484 & 0.121 & \textbf{0.048} \\
{95\% CI (\%)}           & (-0.317, 0.655) & (-0.036, 0.295) & \textbf{(0.001, 0.244)} \\
{VarRed (\%)} & -- & 88.45 & \textbf{93.80} \\
\bottomrule
\end{tabular}
\end{table}

In summary, CUPAC not only delivers stronger variance reduction but also enhances statistical power, providing more precise and timely inference under cluster-based randomization.

%% file: sections/06_conclusion.tex
\section{Conclusion}
We proposed a scalable framework for reliable social A/B testing under network interference.
The framework combines a spillover-contained design stage, enabled by our \emph{Balanced Louvain} clustering algorithm, with a post-experiment analysis stage that recovers statistical efficiency via \emph{CUPAC}-based variance reduction.

Deployed on Kuaishou's production platform, our approach delivers more accurate and reliable treatment-effect estimation in real-world social sharing experiments, substantially improving the assessment of social strategies compared to conventional user-level randomization.
More broadly, the proposed workflow provides a practical paradigm for robust experimentation in industrial-scale social platforms.
Future directions include developing dynamic network-aware randomization that adapts to evolving interaction graphs and exploring hierarchical clustering to support heterogeneous business scenarios across different social applications.

%% file: sections/appendix.tex
\section{Complexity of Balanced Louvain}
\label{appendix:bl_complexity}

We briefly justify that the additional components in Balanced Louvain (soft size penalty and hard post-processing split) do not change the asymptotic runtime of the standard Louvain procedure.

\paragraph{Baseline (Louvain).}
For a graph with $n$ nodes and $m$ edges, each Louvain pass evaluates moving a node only to neighboring communities, so the work per pass is linear in the number of incident edges processed; in practice this yields near-linear scaling in $m$ for sparse graphs.

\paragraph{Soft penalty.}
The soft constraint modifies the node-move score by subtracting $\alpha\,P(|C|)$ (Eq.~\eqref{eq:soft_constraint}). Since $P(|C|)$ depends only on the candidate cluster size, it can be computed and updated in $O(1)$ time per candidate move and therefore adds only constant overhead to each Louvain move evaluation.

\paragraph{Hard split.}
The hard constraint processes only oversized clusters.
For an oversized cluster $C$, we sort its nodes by internal connectivity and move the least-connected nodes to a new cluster until $|C|\le N_{max}$, which costs $O(|C|\log|C|)$ for sorting.
Across the whole partition, each node participates in at most one sort per hard-splitting event of its current cluster, so the total sorting cost is upper bounded by $\sum_C O(|C|\log|C|) \le O(n\log n)$.
Thus, the post-processing stage preserves Louvain's overall asymptotic runtime.

\section{Simulation Study Details}
\label{appendix:simulation}

\subsection{WGSR and $\rho(\mathcal{C})$}

The intra-cluster edge ratio $\rho(\mathcal{C}) = |\{(i,j) \in E : c_i = c_j\}| / |E|$ measures the fraction of edges within clusters at the graph level. WGSR (Equation~\ref{eq:wgsr}) measures spillover containment at the experiment level.

Under cluster-based randomization, WGSR $\geq \rho(\mathcal{C})$ because: (1) edges within the same cluster contribute to both metrics equally; (2) cross-cluster edges between users in the \textit{same experimental group} increase WGSR but not $\rho(\mathcal{C})$. Under UID randomization, WGSR $\approx 0.5$ for balanced experiments.

\subsection{Experimental Setup}

\textbf{Networks.} In order to capture key properties of real social networks (high clustering coefficient, short path lengths) we use Watts-Strogatz small-world networks with $n{=}10{,}000$, $k \in \{4, 10, 20\}$, $p{=}0.1$ for modeling bidirectional friendships.

\textbf{CID Perturbation.} To vary WGSR smoothly:
(1) Apply Louvain clustering;
(2) For perturbation ratio $r \in \{0\%, 11\%, \ldots, 100\%\}$, reset $r{\cdot}n$ users' cluster IDs to their user IDs;
(3) Hash CIDs to 10 groups; Group 1 = treatment (10\%), Group 2 = control (10\%).
This yields WGSR from $\sim$0.95 ($r{=}0\%$) to $\sim$0.50 ($r{=}100\%$).

\textbf{Outcome Model.}
$Y_i = \tau \cdot \mathbf{1}[i \in T] + \delta \cdot \sum_{j \in \mathcal{N}(i)} \mathbf{1}[j \in T] \cdot S_j + \epsilon_i$,
where $\tau{=}1.0$, $\delta{=}0.2$, $S_j \sim \text{Bernoulli}(0.3)$, $\epsilon_i \sim \mathcal{N}(0, 0.01)$.

\subsection{Results and Statistical Validation}

Table~\ref{tab:complete} summarizes results at selected WGSR levels.

\begin{table}[h]
\centering
\caption{Simulation results at selected WGSR levels}
\label{tab:complete}
\small
\begin{tabular}{llccc}
\toprule
Network & WGSR & Obs.\ ATE & 95\% CI & Bias \\
\midrule
\multirow{2}{*}{Low ($\bar{d}{=}4$)}
& 0.95 & 1.215 & [1.214, 1.216] & $-$0.025 \\
& 0.50 & 1.000 & [0.998, 1.002] & $-$0.240 \\
\midrule
\multirow{2}{*}{Medium ($\bar{d}{=}10$)}
& 0.94 & 1.528 & [1.527, 1.530] & $-$0.071 \\
& 0.50 & 1.000 & [0.998, 1.002] & $-$0.600 \\
\midrule
\multirow{2}{*}{High ($\bar{d}{=}20$)}
& 0.94 & 2.044 & [2.042, 2.047] & $-$0.156 \\
& 0.50 & 1.001 & [0.999, 1.003] & $-$1.199 \\
\bottomrule
\end{tabular}
\end{table}

All comparisons between WGSR$\approx$0.95 and WGSR$\approx$0.50 are highly significant ($p < 0.001$, Cohen's $d > 20$). Linear regression of observed ATE on WGSR achieves $R^2 > 0.999$ for all networks, and extrapolation to WGSR$=$1.0 recovers the true ATE within 0.5\%.

\textbf{Practical recommendations:} (1) Compute expected WGSR before experiments to estimate bias reduction; (2) Target WGSR $\geq 0.85$ when spillover is significant; (3) The linear WGSR-ATE relationship enables extrapolation for unbiased estimation.

\textbf{Reproducibility:} 10 WGSR levels $\times$ 30 replications $\times$ 3 networks = 900 experiments. Runtime $\sim$1 min (Apple M1). Code: \url{https://github.com/minxueric/spillover-contained-clustering}

\section{Additional Online Experiment Results}
\label{appendix:online}

\subsection{Engineering implementation for scalability}
\label{appendix:scalability}

Deploying our clustering approach on Kuaishou's large-scale network (hundreds of millions of nodes, billions of edges) requires addressing fundamental scalability limitations of conventional distributed graph systems.
Traditional Spark GraphX \cite{gonzalez2014graphx,zaharia2012resilient,dean2008mapreduce} implementations suffer from (i) massive shuffle overhead during iterative modularity optimization steps across partitions, and (ii) expensive global state synchronization after each iteration.
We overcome these bottlenecks through a re-engineered architecture featuring three core optimizations:

\begin{enumerate}
    \item \textbf{Centralized state management.} We decouple computation from storage by maintaining vertex labels and edge weights in a distributed, in-memory Redis cluster \cite{nishtala2013scaling,redis2023}.
    Compute workers access data through a star-topology pull pattern, eliminating the shuffle phase and drastically reducing network communication costs.

    \item \textbf{Incremental propagation with stable cluster freezing.} We dynamically identify converged macro-communities and freeze their label updates, restricting iterative computation to unstable vertices only.
    This reduces redundant computation while preserving convergence.

    \item \textbf{Adaptive micro-cluster merging via streaming statistics.} We employ T-Digest sketches \cite{dunning2019computing} to compute merging thresholds based on online similarity distribution estimation.
    Multi-threaded workers generate merge candidates in parallel, while a single-threaded arbiter serializes commits, ensuring conflict-free cluster consolidation at scale.
\end{enumerate}

This optimized system enables efficient processing of Kuaishou's billion-edge social graph in production environments, reducing clustering computation time from days to hours while maintaining high-quality community structures.
In our deployment, the end-to-end clustering runtime was reduced from 72 hours to 6 hours.

\subsection{Parallel experiment allocation (CBR vs. UBR)}
\label{appendix:uid-cid-exp}
\begin{figure}[h]
    \centering
    \includegraphics[width=0.85\linewidth]{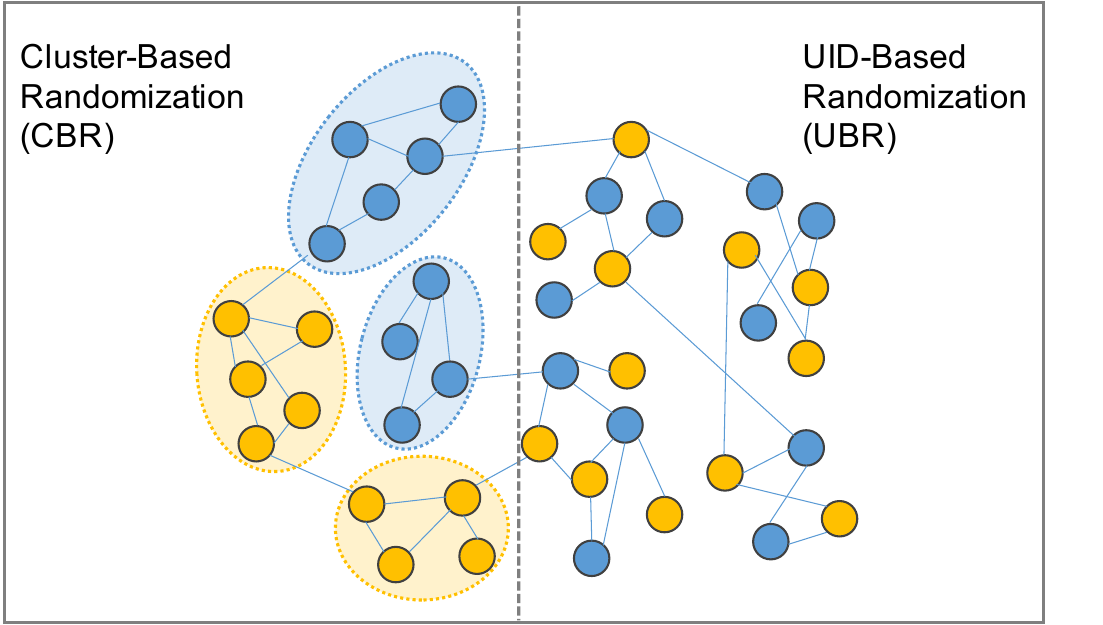}
    \caption{Illustration of our joint \texttt{cluster-based} and \texttt{uid-based} experimental design.
    We first allocate part of the traffic according to CBR to contain spillover.
    The remaining traffic is then assigned using UBR, enabling a parallel comparison.}
    \label{fig:uid-cid-exp}
\end{figure}

Figure~\ref{fig:uid-cid-exp} illustrates how we run CBR and UBR in parallel without cross-contamination.
We first divert a fixed portion of users into the cluster-based randomization (CBR) experiment, where the unit of assignment is the cluster ID (CID).
The remaining users are then independently assigned under user-based randomization (UBR) at the user ID (UID) level.
This setup ensures the two experiments are mutually exclusive, enabling a clean side-by-side comparison of the same strategy under different randomization schemes.

\subsection{Retention lift under different randomization schemes}
\label{appendix:lt7}
\begin{figure}[h]
    \centering
    \includegraphics[width=0.8\linewidth]{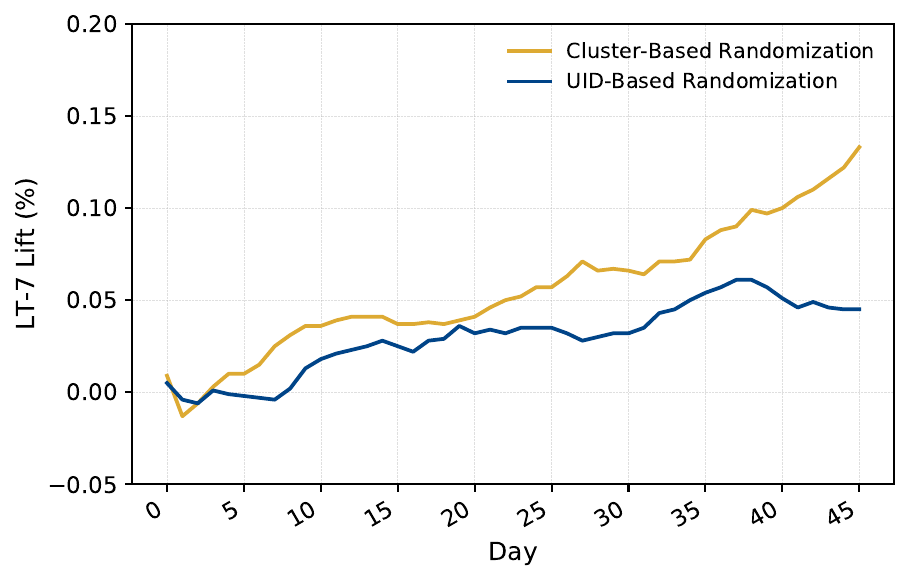}
    \caption{LT-7 lift comparison of cluster-based (CBR) vs. user-based (UBR) randomization under the sharing strategy.}
    \label{fig:LT7}
\end{figure}

Figure~\ref{fig:LT7} reports the day-by-day lift in 7-day retention (LT-7) for the same sharing strategy under two randomization schemes: user-based randomization (UBR) and our spillover-contained cluster-based randomization (CBR).
Consistent with the discussion in Section~\ref{sec:online_results}, CBR yields a clearer and statistically significant retention gain, whereas UBR is largely indistinguishable from noise, highlighting how spillover can mask downstream effects under user-level assignment.

\section{Detailed Clustering Ablation Experiments}
\label{sec:appendix_clustering}

\subsection{LPA with Large Cluster Constraints}
\label{sec:appendix_lpa}

We extend standard Label Propagation Algorithm (LPA) with large cluster constraints as a baseline. The key modifications are:

\begin{algorithm}[h]
\caption{LPA with Large Cluster Constraints}
\label{alg:lpa_constrained}
\begin{algorithmic}[1]
\Require Graph $G = (V, E, W)$, large cluster threshold $\theta$
\Ensure Cluster assignment $\ell: V \rightarrow \mathcal{L}$

\State Initialize: $\ell(i) \leftarrow i$ for all $i \in V$; $\mathcal{L}_{large} \leftarrow \emptyset$

\Statex \textbf{Phase 1: Propagation with Large Cluster Filtering}
\Repeat
    \For{each node $i \in V$}
        \If{$\ell(i) \notin \mathcal{L}_{large}$}
            \State $\ell(i) \leftarrow \arg\max_{l \notin \mathcal{L}_{large}} \sum_{j \in N(i), \ell(j)=l} w_{ij}$
        \EndIf
    \EndFor
    \State Update $\mathcal{L}_{large} \leftarrow \{l : |C_l| > \theta\}$
\Until{convergence}

\Statex \textbf{Phase 2: Large Cluster Re-propagation}
\State $V_{large} \leftarrow \{i : \ell(i) \in \mathcal{L}_{large}\}$
\State Reset: $\ell(i) \leftarrow i$ for all $i \in V_{large}$
\Repeat
    \For{each $i \in V_{large}$}
        \State Propagate among $V_{large}$ neighbors, excluding $\mathcal{L}_{large}$
    \EndFor
\Until{convergence}
\State \Return $\ell$
\end{algorithmic}
\end{algorithm}

This heuristic approach lacks an optimization objective (unlike modularity in Louvain), resulting in lower clustering quality as shown in our experiments.

\subsection{Parameter Sensitivity}

\paragraph{Balance Factor $\alpha$}
Table~\ref{tab:alpha_sensitivity} shows the effect of varying $\alpha$ using soft constraint only. A negative $N_{max}$ value disables the post-processing hard constraint, isolating the effect of the soft constraint (see Section~\ref{sec:balanced_louvain} for details).

\begin{table}[h]
\centering
\small
\caption{Sensitivity to balance factor $\alpha$ (soft constraint only, $N_{max}=-40,000$).}
\label{tab:alpha_sensitivity}
\begin{tabular}{ccccc}
\toprule
{$\alpha$} & {Modularity} & {Intra-edge} & {Variance} & {Max Cluster} \\
\midrule
0 (baseline) & \textbf{0.648} & \textbf{0.548} & 1,839,930 & 109,296 \\
0.2 & 0.631 & 0.506 & 448,563 & 51,390 \\
0.3 & 0.623 & 0.511 & 360,540 & 45,722 \\
0.5 & 0.607 & 0.512 & \textbf{358,535} & \textbf{38,365} \\
\bottomrule
\end{tabular}
\end{table}

As $\alpha$ increases, the soft constraint more effectively limits cluster sizes: max cluster decreases from 109K to 38K, and variance drops significantly from 1.84M to 359K. This comes at the cost of reduced modularity (0.648 $\to$ 0.607). We recommend $\alpha \in [0.2, 0.3]$ for balanced performance.

\paragraph{Maximum Cluster Size $N_{max}$}
Table~\ref{tab:nmax_sensitivity} shows the effect of varying $N_{max}$ with fixed $\alpha=0.5$.

\begin{table}[h]
\centering
\small
\caption{Sensitivity to maximum cluster size $N_{max}$ ($\alpha=0.5$).}
\label{tab:nmax_sensitivity}
\begin{tabular}{cccccc}
\toprule
{$N_{max}$} & {Modularity} & {Intra-edge} & {Variance} & {Max Cluster} & {Max/N$_{max}$} \\
\midrule
30,000 & 0.613 & 0.501 & \textbf{264,415} & 29,998 & 100\% \\
40,000 & 0.624 & 0.500 & 357,992 & 34,484 & 86\% \\
50,000 & 0.621 & \textbf{0.518} & 461,524 & 45,791 & 92\% \\
\bottomrule
\end{tabular}
\end{table}

The algorithm consistently controls max cluster size below $N_{max}$. Tighter $N_{max}$ leads to lower variance but may reduce intra-edge ratio.

\subsection{Comparison with Resolution Parameter}

Table~\ref{tab:resolution_appendix} compares our method with varying resolution parameter $\gamma$.

\begin{table}[h]
\centering
\small
\setlength{\tabcolsep}{3pt}
\caption{Comparison with resolution-based cluster size control.}
\label{tab:resolution_appendix}
\resizebox{\columnwidth}{!}{%
\begin{tabular}{lcccccc}
\toprule
{Method} & {$\gamma$} & {Modularity} & {Intra-edge} & {Variance} & {Max Cluster} & {Ctrl} \\
\midrule
Louvain & 0.5 & 0.640 & \textbf{0.557} & 1,855,385 & 100,511 & $\times$ \\
Louvain & 1.0 & \textbf{0.648} & 0.548 & 1,839,930 & 109,296 & $\times$ \\
Louvain & 1.5 & 0.646 & 0.520 & 788,000 & 76,173 & $\times$ \\
Louvain & 2.0 & 0.646 & 0.500 & 387,911 & 50,202 & $\times$ \\
\midrule
Ours & 1.0 & 0.622 & 0.511 & \textbf{373,842} & 36,877 & \checkmark \\
\bottomrule
\end{tabular}%
}
\end{table}

Resolution-based approaches cannot precisely control maximum cluster size and have high variance at low $\gamma$. Even at $\gamma=2$, variance (388K) is comparable to ours (374K), but our method provides precise control and higher intra-edge ratio (0.511 vs 0.500).

\section{Variance Reduction}
\label{sec:appendB}

\begin{table}[H]
\centering
\caption{Variance reduction (\%) relative to DIM for CUPED and CUPAC across LT-7 metrics}
\label{tab:variance_reduction}
\begin{tabular}{lcc}
\toprule
 & {CUPED (\%)} & \textbf{CUPAC (ours) (\%)} \\
\midrule
Kuaishou LT-7 & 85.95 & \textbf{92.45 (+6.50pp)} \\
Bi LT-7       & 89.11 & \textbf{93.77 (+4.66pp)} \\
\bottomrule
\end{tabular}
\end{table}

We evaluated two LT-7 metrics over the 45-day experimental period: Kuaishou LT-7, measured on the main Kuaishou app only, and Bi LT-7 (Kuaishou main app and its Lite version).
Table~\ref{tab:variance_reduction} reports the variance reduction of CUPED and CUPAC relative to the DIM baseline across LT-7 metrics. Results are averaged over the 45-day experimental window. For both Kuaishou LT-7 and Bi LT-7, CUPAC achieves higher variance reduction than CUPED, with gains of 6.50pp and 4.66pp respectively.